\documentclass[12pt]{article}
\usepackage{graphics}
\usepackage{amsmath,amsfonts,amssymb,graphicx}
\begin{document}
\pagestyle{plain} \baselineskip 0.4cm \vspace{0.1cm}
\begin{center}
{\huge \bf Multiquark Cluster Form Factors In the Relativistic
Harmonic Oscillator Model} $^{*}$ \footnote[1]{{The work was
supported in part by the National Natural Science Foundation of China
(11365002), Guangxi Natural Science Foundation for Young Researchers
(2013GXNSFBB053007, 2011GXNSFA018140), Guangxi Education Department
(2013ZD049), Guangxi Grant for Excellent Researchers (2011-54), and the
 Guangxi University of Science and Technology Foundation for PhDs
(11Z16).}}
\end{center}

\vspace{1.2 cm}
\begin{center}
{\bf Wu Qing$^{1}$, Zhou Li-juan$^{2}$, Xiang Qian-fei $^{3}$, and
Ma Wei-xing$^{3}$}
\\

\vspace{0.4cm} ${1}$ Department of Physics, School of Science,
Qingdao University, QingDao, 266076, China
\\

\vspace{0.4cm} $^{2}$ School of Science, Guangxi University of
Science and Technology, LiuZhou, 545006, China
\\

\vspace{0.4cm}
$^{3}$ Institute of High Energy Physics, Chinese
Academy of Sciences, Beijing, 100049, China

\end{center}

\vspace{0.4cm}
\begin{center}
{\Large \bf Abstract}
\end{center}

A QCD multiquark cluster system is studied in the relativistic harmonic oscillator potential model
(RHOPM), and the electromagnetic form factors of the pion, proton and deuteron in the RHOPM are predicted.
The calculated theoretical results are then compared with
existing experimental data, finding very good agreement between
the theoretical predictions and experimental data for
these three target particles. We claim that this model can be applied to study
QCD hadronic properties, particularly neutron properties, and to
find six-quark cluster and/or nine-quark cluster probabilities in
light nuclei such as helium $^{3}He$ and tritium $^{3}H$.
This is a problem of particular importance and interest in quark
nuclear physics.

\vspace{1.0cm} \noindent {\bf Key words:}  multiquark
cluster system form factors, Relativistical Harmonic Oscillator Potential Model,
quarks, QCD.

\vspace{0.4cm} \noindent {\bf PACS Numbers: 24.85.+p,12.38.Lg,
12.38.Mh.}

\section{Introduction}

Hadrons are particles which interact by the strong interaction.
Hadrons have two classification: mesons and baryons. Mesons, with
meson number $|M>=\frac{1}{\sqrt{3}}|q_{\alpha}\bar{q}_{\beta}>$,
are made up of a quark and an antiquark, with q denotes it's quark
state and $\alpha$ is the quantum numbers. Baryons, with baryon
number $|B>=
\frac{1}{\sqrt{6}}\epsilon^{\alpha\beta\gamma}|q_{\alpha}q_{\beta}q_{\gamma}>$,
are made up of three quarks. Apart from these, there is much more
picture than this, the constituent quarks being surrounded by a
cloud of gluons, the exchange particles for the strong force${[1]}$.

The strong force hold two or more quarks together, which formed
hadron. Valence quarks determine the quantum numbers of hadrons,
besides these, any hadron is made up of an indefinite number of sea
quarks, antiquarks and gluons, which do not influence the quantum
numbers of hadrons. Here, we investigate only valence quark cluster
systems and do not consider the existence of sea quarks and gluons.

Our present understanding of hadrons as extended objects
containing colored quarks and gluons suggests that a nucleus might
not always behave as a simple collection of nucleons. Even in the loosely
bound deuteron there is a few percent probability that the nucleons
are separated by a distance less than their radius. In such a
situation it seems reasonable that instead of talking of two
clusters of three quarks one should speak of a single six-quark
system${[2]}$. Of course, if we were to decompose the six-quark
system into clusters they could be either color singlets or
octets${[3]}$. A specific estimate of about $7\%$ is obtained by
theoretical models for the deuteron form factor${[4]}$.

In short, strongly interacting composite
particles can be viewed as multiquark clusters. The deuteron is thus made
up of six quarks if the proton and neutron are overlapping; in the same
circumstances, $^{3}He$ is a system of nine quarks.

Many studies have been done using conventional methods for the form factors of strong interaction
composite particles. More recently, however,
Refs.[5,6] study hadron form factors in perturbative QCD and QCD-inspired models.
We work in the framework of a relative harmonic oscillator potential model (RHOPM),
 an N-valence quark cluster system where the quarks move
in a relativistic harmonic oscillator potential.

\section{Form factors of multiquark bound states}

Closely following Ref.[7], we consider a system consisting of N
quarks moving in the field of a relativistic harmonic oscillator
potential. The wave function has the form
\begin{eqnarray}
\Psi^{N}_{P}( x_{1},x_{2}, \dots, x_{N})=
\tilde{A}\Phi_{N}(x_{1},x_{2},\dots, x_{N}) U^{N} (\vec{P})
\end{eqnarray}
where $\tilde{A}$ is the quark antisymmetrization operator,
$\Phi_{N}(x_{1},x_{2}, \dots, x_{N})$ is the space-time wave
function, and $U^{N}(\vec{P})$ is the spin wave function. We assume
that the wave function $\Phi_{N}$ obeys the Klein-Gordon equation
within a relativistic harmonic oscillator potential${[7]}$
\begin{eqnarray}
\{\sum_{i=1}^{N}p^{2}_{i} +
\kappa^{2}[\sum_{i>j}^{N}\sum_{j=1}^{N-1}(x_{i} -
x_{j})^{2}]\}\Phi_{N}(x_{1},x_{2}, \dots, x_{N}) = 0
\end{eqnarray}
where $p_{i}= -i\partial/\partial x_{i}$ and $\kappa$ are,
respectively, the 4-momentum and the oscillator parameter, $x_{i}$
is the 4-coordinate of the $i$-th quark. Let us assume isospin
invariance and all quark masses are equal. After some derivation,
one can represent Eq. (2) in the form
\begin{eqnarray}
(P^{2}- M^{2}_{p})\Phi_{Nq}(r_{0}, r_{1}, \dots, r_{N-1},P)= 0,
\end{eqnarray}
\begin{eqnarray}
M_{p}^{2}= -2\alpha_{N}a^{+}_{i\mu}a_{i\mu} + const,
\end{eqnarray}
\begin{eqnarray}
\alpha_{N} =\kappa N \sqrt{N}
\end{eqnarray}
where $P$ is the total momentum, $M_{P}$ is the mass of the system,
$a^{+}_{i\mu}$ is particle creation operators, and $a_{i\mu}$ is
particle annihilation operators. With the Takabayashi
condition${[8]}$, removing nonphysical oscillations, $p^{\mu}
a^{+}_{i\mu}\Phi_{Nq}=0$, one gets the solution
\begin{eqnarray}
\Phi_{Nq}(r_{0},r_{1}, \dots, r_{N-1}, P) =(\frac{\alpha_{N}}{\pi
N})^{N-1}exp(\frac{\alpha_{N}}{2N}K^{\mu\nu}\sum_{i=1}^{N-1}r_{i\mu}r_{i\nu}),
\end{eqnarray}
and
\begin{eqnarray}
\Phi_{N}(x_{1},x_{2},\dots, x_{N}) =
exp[ip_{\mu}X_{\mu}]\Phi_{Nq}(r_{0}, r_{1}, \dots, r_{n-1}, P)
\end{eqnarray}
where $K^{\mu\nu}=g^{\mu\nu}-2p^{\mu}p^{\nu}/P^{2}$. From Eq. (6),
then, the N-quark cluster wave function $\Phi_{Nq}$, with the
subsidiary condition formulated by Takabayasi condition, can be
written explicitly ($n = N-1$) as:
\begin{eqnarray}
\Phi_{Nq}= (\frac{\alpha_{N}}{\pi
N})^{n}exp[\frac{\alpha_{N}}{2N}(g^{\mu\nu}-2\frac{p^{\mu}p^{\nu}}{M^{2}_{Nq}})
(\sum^{n}_{i=1}r^{i}_{\mu}r^{i}_{\nu})],
\end{eqnarray}
where the plane wave part for the center of mass
coordinate has been dropped. It is well known that the wave function $\Phi_{Nq}$in
Eq.~(8) is characterized by the Lorentz contraction effect.

Transforming the non-relativistic spin wave function in the rest
frame. Then the spin wave function $U^{N}(\vec{P})$ can be given
by${[9]}$
\begin{eqnarray}
U^{N}(\vec{P}) = B (\vec{P})U^{N}(0),
\end{eqnarray}
\[U^{N}(0)= \left(
 \begin{array}{c}\chi\\  0
 \end{array}
 \right)\],
 \begin{eqnarray}
 B(\vec{P}) = exp[\frac {b}{2|\vec{P}|}\rho_{1}(\vec{P} \cdot \vec {\sigma })]=
 exp[\rho_{1} b H ],
 \end{eqnarray}
\[
\rho_{1}=\left(\begin{array}{cc}
0 & 1 \\
1 & 0
\end{array}
\right)
\]
where $\chi$ is the non-relativistic spin function, $H = (\vec{P}
\cdot \vec {\sigma})/2|\vec{P}|$, $b = cosh^{-1}p_{0}/M_{P}$ and
$\vec{\sigma} = \sum_{i=1}^{N} \vec {\sigma}^{i}$, with
$\vec{\sigma}^{i}$ is the Pauli matrices for the $i$-th quark.

According to Ref.[7], the electromagnetic action can be written as
\begin{eqnarray}
I_{em} =\int \prod_{i=1}^{N}dx_{i}\sum_{k}^{N}j_{k\mu}(x_{1},x_{2},
\dots, x_{N}) A_{\mu}(x_{k}) \equiv \int dX J^{N}_{\mu}(X)A_{\mu
}(X),
\end{eqnarray}
with
\begin{eqnarray}
j_{k\mu}(x_{1},x_{2},\dots, x_{N})=-i\Psi^{N}_{p'} N
e_{k}[g_{E}(q^{2})\frac{\overrightarrow {\partial}}{\partial x_{k
\mu}} + ig_{M}(q^{2})
\sigma^{k}_{\mu\nu}(\frac{\overrightarrow{\partial}}{\partial x_{k
\nu}}+ \frac{\overleftarrow{\partial}}{\partial x_{k
\nu}})]\Psi^{N}_{p}.
\end{eqnarray}
In Eq. (12), $\Psi^{N}_{p,(p')}$ is the initial (final) wave
function as given in Eq.(1), $e_{k}$ and $\sigma^{k}_{\mu\nu}$ are,
respectively, the charge and the spin matrices of the $k$-th quark,
and $\sigma^{k}_{ij}=\varepsilon_{ijl} \sigma^{k}_{l}$,
$\sigma^{k}_{i4}=\sigma^{k}_{4i}=\rho_{1}\sigma^{k}_{i}$.
$g_{E}(q^{2})$ and $g_{M}(q^{2})$ are quark charge and magnetic form
factors, and $q = p'-p$ . Putting the wave function $\Phi_{N}$ from
Eq. (7) into Eqs. (11) and (12), and computing the integrals, the
matrix elements of the effective current, $J^{N}_{\mu}(0)$ can be
given as:
\begin{eqnarray}
\langle p's'\mid J^{N}_{\mu}(0) \mid  p s \rangle =
\frac{I^{N}(q^{2})}{\sqrt{2p_{0}p'_{0}}}\sum_{k=1}^{N}(\bar{U}^{N}_{s'}(p')
\Gamma_{k,\mu}U^{N}_{s}(p)),
\end{eqnarray}
where
\begin{eqnarray}
\Gamma_{k,\mu}= e_{k}[( p_{\mu}+ p'_{\mu})I_{N}(q^{2})
g_{E}(q^{2})-iNg_{M}(q^{2})\sigma^{k}_{\mu\nu}q_{\nu}].
\end{eqnarray}
Where $I^{N}(q^{2})$ and $I_{N}(q^{2})$ have the form:
\begin{eqnarray}
I^{N}(q^{2}) = \frac{1}{( 1 + q^{2}/2M_{Nq}^{2})^{N-1}}
exp[-\frac{N-1}{4\alpha_{N}}(\frac{q^{2}}{1 + q^{2}/2 M_{Nq}^{2}})],
\end{eqnarray}

\begin{eqnarray}
I_{N}(q^{2})= \frac{1+ Nq^{2}/2M^{2}_{Nq}}{1+ q^{2}/2M^{2}_{Nq}}.
\end{eqnarray}

We now study the N-quark bound state and write down it's form factor
\begin{eqnarray}
F_{Nq}(q^{2})= \int \Phi_{Nq}^{*}(r^{1},\dots ; P_{F})exp[-iq
\sum_{i=1}^{n} u_{1}^{i}r^{i}] \Phi_{Nq}(r^{1},\dots ; P_{I})\times
d^{4}r^{1}\dots d^{4}r^{n},
\end{eqnarray}
where $u^{i}_{1}$ is the first component of eigenvector $\bf
{u}^{i}$, obey the normalization condition
\begin{eqnarray}
\sum_{i=1}^{n}|u_{1}^{i}|^{2}= \frac{n}{N},
\end{eqnarray}
where $n=N-1$. After some derivation using the operator
$a^{i}_{r\mu} =\frac{1}{\sqrt{2\alpha_{N}}}(\sqrt{N} p^{i}_{r\mu}
-i\frac{\alpha_{N}}{\sqrt{N}}r^{i}_{\mu})$, where
$\alpha_{N}=N^{3/2}k$, and Eq. (17) takes the form

\begin{eqnarray}
F_{Nq}(Q^{2}) = \frac{1}{[1+(Q^{2}/2M_{Nq}^{2})]^{n}}
exp[ -\frac{n}{4\alpha_{N}}\frac {Q^{2}}{1+ (Q^{2}/2M^{2}_{Nq})}],
\end{eqnarray}
where $Q^{2}=-q^{2}$.

For the two-quark cluster of the pion ($N = 2$),
the form factor can be written ($n = N-1=1$) as

\begin{eqnarray}
 F_{\pi}(Q^{2})=\left[1+\frac{Q^{2}}{2M_{\pi}^{2}}\right]^{-1}
 \exp\left[-\frac{1}{4\alpha_{\pi}}\frac{Q^{2}}{1+\frac{Q^{2}}{2M_{\pi}^{2}}}\right].
\end{eqnarray}

Similarly, for the nucleon three-quark cluster ($N=3$), the form factor
can be expressed as

\begin{eqnarray}
 F_{N}(Q^{2})=\left[1+\frac{Q^{2}}{2M_{N}^{2}}\right]^{-2}
 \exp\left[-\frac{1}{2\alpha_{N}}\frac{Q^{2}}{1+\frac{Q^{2}}{2M_{N}^{2}}}\right].
\end{eqnarray}

For the six-quark cluster system of deuteron, once the distance
between the proton and the neutron is less than their radius, the
form factor of six-quark cluster can be given as
\begin{eqnarray}
F_{D}(Q^{2}) = \frac{1}{[1+(Q^{2}/2M_{D}^{2})]^{5}} exp[
-\frac{5}{4\alpha_{D}}\frac {Q^{2}}{1+ (Q^{2}/2M^{2}_{D})}].
\end{eqnarray}

In Section 3, we compare our present theoretical results for the proton, pion and
 deuteron form factors with the experimental data.

\section{Comparison with experimental data}

Fig.~1 and Fig.~2 show our present calculated
electromagnetic form factors for the proton and pion respectively, compared with the
corresponding experimental data${[10-12]}$. As Figs. 1 and 2 show, there is very good
agreement between the theoretical and experimental data.

Fig. 3 shows the fit to the deuteron scalar form factor $A(Q^{2})$
at high energies, where it should be possible to predict the six-quark
cluster probability in the deuteron wave function. Our
predicted result for the deuteron electromagnetic form factor
$F_{6q}(Q^{2})sin^{2}(\theta)$ is approximately identical to the
deuteron scalar form factor $A(Q^{2})$ of the Rosenbluth separation at high energies, where
$sin^{2}(\theta)$ is the probability of the six-quark cluster component in the deuteron.
Therefore, comparing the theoretically calculated result $F_{6q}(Q^{2})sin^{2}(\theta)$ with the
experimental data for $A(Q^{2})$, we can get the value of $sin^{2}(\theta)$. This is an interesting and important
issue in modern nuclear physics and hadron physics.

\begin{figure}[hb]
\begin{center}
\includegraphics[height=8 cm, width=10 cm]{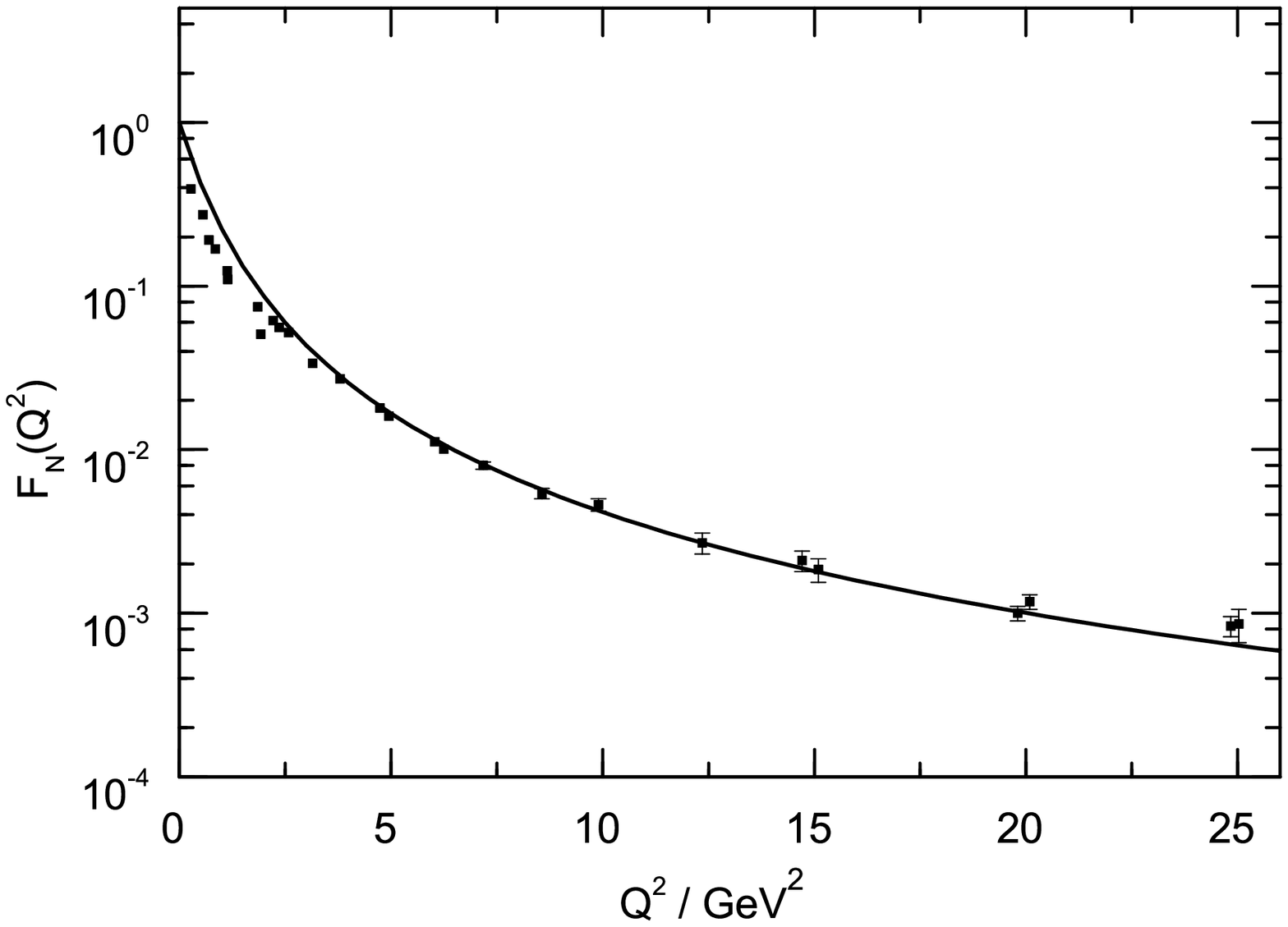}
\caption{$Q^{2}$-dependence of nucleon form factor $F_{N}(Q^{2})$
and comparison with experimental data from Ref.[10].} \label{Fig1}
\end{center}
\end{figure}

\begin{figure}
\begin{center}
\includegraphics[height=8 cm,width=10.cm]{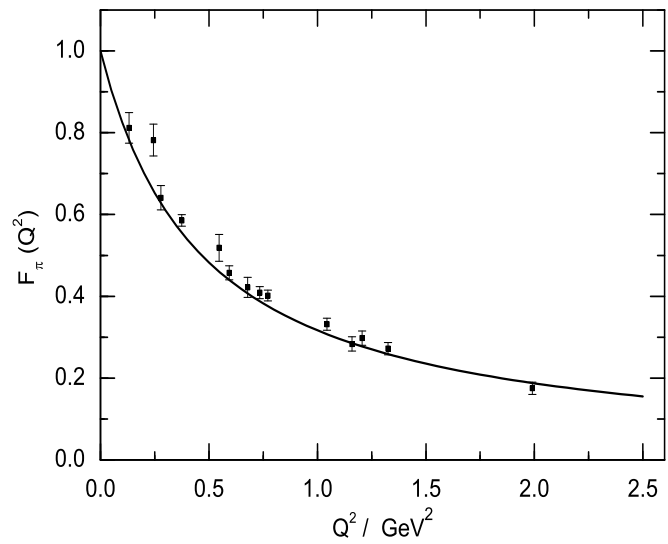}
\caption{$Q^{2}$-dependence of pion form factor $F_{\pi}(Q^{2})$ and
comparison with experimental data from Ref.[11].} \label{Fig2}
\end{center}
\end{figure}

\begin{figure}
\begin{center}
\includegraphics[height=8 cm,width=10.cm]{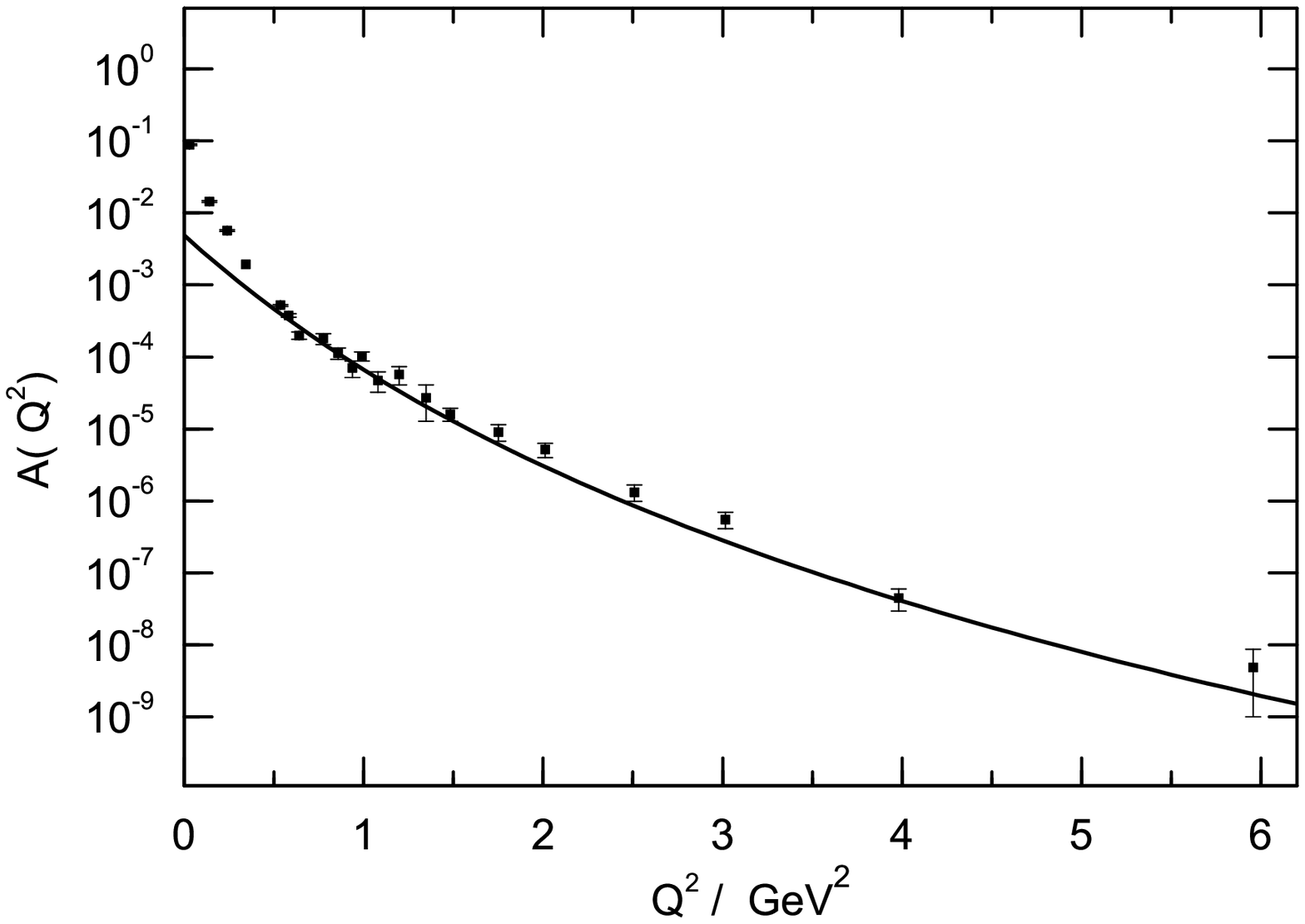}
\caption{$Q^{2}$-dependence of deuteron form factor $F_{D}(Q^{2})$
and comparison with experimental data from Ref.[12].} \label{Fig3}
\end{center}
\end{figure}

\section{Conclusions}
In this paper, we studied the electromagnetic
form factors of multiquark clusters in the RHOPM. Based on the belief
that strongly interacting composite particles are made up of valence quarks,
and assuming quarks move individually within the relativistic harmonic
oscillator potential, we have calculated the electromagnetic form factor of the proton, the pion
and the deuteron in the RHOPM. This model gives a fairly good simple description
of these three particle structures provided only one arbitrary parameter,
$g_{E}(q^{2})=1.0$, is applied. Agreement with the corresponding experimental
data is very good for all three particles.

The study of electromagnetic form factors of hadrons and nuclei has been a longstanding
physical problem, on which much research work has already been published ${[13]}$.
However, our theoretical investigations give not only a simple analytical expression of
electromagnetic form factors of multiquark cluster systems, which is very useful
for practical investigations, but also the results have pointed out a way to find
six-quark and nine-quark cluster probabilities in nuclei. For example, comparing our
calculated form factor for the deuteron $F_{6q}sin^{2}(\theta)$ at high energies with
the Rosenbluth separation form factor $A(Q^{2})$ gives the deuteron six-quark cluster probability, $sin^{2}(\theta)$,
to be about $7\%$, since $F_{D}(Q^{2}) = F_{np}(Q^{2})cos^{2}(\theta) +
F_{6q}(Q^{2})sin^{2}(\theta)$ and at high energies $F_{6q}(Q^{2})sin^{2}(\theta) = A(Q^{2})$.
This model can easily be extended to other mesons and baryons, as
well as any system with a number of quarks larger than three, e.g. light nuclei such as $^{3}He$
and $^{3}H$. Needless to say, finding six-quark and /or nine-quark probabilities in many-body
nucleon systems is an important and interesting issue in nuclear physics which is helpful for the development of
quark nuclear physics.

\end{document}